\documentclass[aps,prd,twocolumn,preprintnumbers,nofootinbib,superscriptaddress,amsmath]{revtex4-2}
\usepackage{graphicx}
\usepackage{amsmath}
\usepackage{amssymb}
\usepackage{amsfonts}
\usepackage{hyperref}
\usepackage{url}
\usepackage{color}
\usepackage{bm}
\usepackage{array}

\newcommand{\be}{\begin{equation}}
\newcommand{\ee}{\end{equation}}

\newcommand{\ba}{\begin{eqnarray}}
\newcommand{\ea}{\end{eqnarray}}
\newcommand{\bea}{\begin{eqnarray}}
\newcommand{\eea}{\end{eqnarray}}
\newcommand{\nn}{\nonumber}

\renewcommand{\[}{\begin{equation}}
\renewcommand{\]}{\end{equation}}

\usepackage{array}
\makeatother

\begin{document}

\preprint{IFT-UAM/CSIC-23-83}

\title{Effects of orbital precession on hyperbolic encounters}

\author{Marienza Caldarola}
\email{marienza.caldarola@csic.es}
\affiliation{Instituto de F\'isica Te\'orica UAM-CSIC, Universidad Aut\'onoma de Madrid,
Cantoblanco, 28049 Madrid, Spain}

\author{Sachiko Kuroyanagi}
\email{sachiko.kuroyanagi@csic.es}
\affiliation{Instituto de F\'isica Te\'orica UAM-CSIC, Universidad Aut\'onoma de Madrid,
Cantoblanco, 28049 Madrid, Spain}
\affiliation{Department of Physics and Astrophysics, Nagoya University, Nagoya, 464-8602, Japan}

\author{Savvas Nesseris}
\email{savvas.nesseris@csic.es}
\affiliation{Instituto de F\'isica Te\'orica UAM-CSIC, Universidad Aut\'onoma de Madrid,
Cantoblanco, 28049 Madrid, Spain}

\author{Juan Garcia-Bellido}
\email{juan.garciabellido@uam.es}
\affiliation{Instituto de F\'isica Te\'orica UAM-CSIC, Universidad Aut\'onoma de Madrid,
Cantoblanco, 28049 Madrid, Spain}

\date{\today}

\begin{abstract}
The hyperbolic encounters of two massive objects are characterized by the emission of a gravitational wave burst, with most of the energy released during the closest approach (near the periastron).
The detection of such events, different from the well-known inspiral emission, would be an interesting discovery and provide complementary information to observations of binary mergers of black holes and neutron stars in the observable Universe, shedding light, for instance, on the clustering properties of black holes and providing valuable hints on their formation scenario.
Here, we analyze the dynamics of such phenomena in the simplest case where two compact objects follow unbound/hyperbolic orbits. Moreover, we explore the effects of orbital precession on the gravitational wave emission, since the precession encodes certain general relativistic effects between two bodies. We also provide templates for the strain of gravitational waves and the power spectrum for the emission, and analytical expressions for the memory effect associated with such signals.
\end{abstract}

\maketitle

% -------------------
\section{Introduction \label{sec:intro}}
The detection of gravitational waves (GWs) has been one of the most remarkable scientific results of the past decade. The first detection of GWs by the LIGO-Virgo collaboration dates back to 2015~\cite{Abbott:2016blz}, and it has opened a new era in the multimessenger astronomy and a new window into the understanding of the Universe. The interaction of GWs with matter is extremely weak; this has made direct GW detection extremely challenging, but current detectors have already reached such sensitivity, and ongoing studies on this topic are leading to further improvements. However, a positive aspect about the weakness of the gravitational interaction is that GWs could travel over cosmological distances, allowing us to obtain a more precise scenario of their sources and providing a unique way to study phenomena such as black hole (BH) and neutron star (NS) mergers.

In the last few years, the improved sensitivity of Advanced LIGO~\cite{LIGOScientific:2014pky} and Advanced Virgo~\cite{VIRGO:2014yos} detectors has led to an increased number of detected events, including a number of BH mergers and the detection of two NS inspirals (GW170817, GW190425) \cite{TheLIGOScientific:2017qsa} \cite{Abbott_2020}.
With progress in detecting interactions of binary systems and the success of GW astronomy~\cite{TheLIGOScientific:2017qsa,KAGRA:2013rdx}, there has been a growing interest in studying scattering events related to GW astronomy~\cite{Chatterjee2017,Cho_2018,Kocsis:2006hq,OLeary:2008myb}, as this could be useful to deepen and reveal characteristics of astrophysical sources and properties of galaxies and clusters.

In the context of GW emission, a possible scenario is the emission of gravitational radiation by massive bodies that move on unbound orbits, emitting gravitational Bremsstrahlung radiation~\cite{Turner:1978ApJ,Turner:1977tm}. A hyperbolic encounter refers to a system where two objects, such as BHs or NSs, pass by each other in a hyperbolic trajectory, meaning that the objects approach each other from a large distance, interact for a short time and then proceed on separate paths. In literature, there are already studies about parabolic~\cite{OLeary:2008myb} and hyperbolic encounters~\cite{Capozziello:2008ra, DeVittori:2012da,Grobner:2020fnb}, respectively, assuming a Keplerian orbit~\cite{Peters:1963,Peters:1964}. In particular, hyperbolic encounters are expected to take place in dense BH clusters~\cite{Trashorras:2020mwn}, where a fraction of them is expected to be not in a bounded system and then BHs gravitationally scatter each other, emitting GWs. This happens if the velocity or the relative distance between the two bodies is large to prevent a BH capture. They can produce interesting consequences, such as spin induction~\cite{Jaraba:2021ces, Nelson:2019czq}, subsequent mergers~\cite{Healy:2009zm}, the generation of a stochastic GW background~\cite{Garcia-Bellido:2021jlq}, the possibility of exploring dynamical friction from dark matter~\cite{Chowdhuri:2023}, etc.

The gravitational waveform in the time domain results in a burst-like signal with a characteristic frequency peak~\cite{Garcia-Bellido:2017qal,Garcia-Bellido:2017knh}, which occurs due to the rapid changes in the gravitational field as the objects approach and interact rapidly during the close encounter. This burst is a short-duration signal: an appropriate modeling is important for its interpretation, to be distinguished and properly recognized by future GW detectors~\cite{Kocsis:2006hq,Morras:2021atg,Mukherjee:2020hnm,Dandapat_2023}, such as Einstein Telescope~\cite{Maggiore:2020JCAP,Punturo2010CQGra} and Cosmic Explorer~\cite{Evans:2021}. However, it should be noted that events of this type are very tiny, so they would be easily hidden by noise in the detectors. The features of the emitted GW peak signal depend on different parameters, such as the total mass of the system, relative velocities, and relative orientations of the objects involved. Therefore, by analyzing the burst waveform, it may be possible to shed light on the properties of objects and the dynamics of the encounter, providing valuable information about astrophysical objects and contributing to the understanding of the Universe and gravitational phenomena.

In this paper, we aim to provide a comprehensive theoretical overview of the formalism underlying GW emission during hyperbolic encounters between two massive objects and discuss how this is affected by the precession of the orbit. The precession could lead, for instance, to a deviation from original trajectory or a change in orbital velocity. Although this effect is incredibly small, it reflects the relativistic effect between two bodies and it might be susceptible to future observation. 

In order to consider the precession of an orbit, we could focus on two main approaches, namely using the post-Newtonian (PN) corrections formalism, then expanding relevant quantities as a power series in terms of $v/c$ (see, e.g.,~\cite{DeVittori:2014psa,Cho_2018}), or directly incorporating the relevant corresponding PN corrections into the expression of the trajectory. In the latter case, the expression of the trajectory is directly modified accounting for the relativistic effects responsible for precession. This approach can be more straightforward and is what we will follow in our analysis. However, it might involve more complex calculations, as we will show for the determination of the power spectrum.

Our purpose is to offer a first fully analytic estimate of consequences of orbital precession on strain and power emitted in GWs by two non-spinning compact objects on hyperbolic encounter. Considering precession is significant to assess gravitational effects and properly studying the physics of such encounters, since this might have significant implications for the dynamics of orbits and analysis of GWs signals. Therefore, in this paper, we also provide considerations about the validity of our approximation, since in the limit of very close encounter other effects should be considered, which need a different description.

This paper is organized as follows. In Sec. \ref{SectionII}, we review the basics of the theory without considering the effect of the precession, deriving the main relations determining the geometry and physics of hyperbolic encounters. We report analytical expressions for the GW strain amplitude and power spectrum for the emission. We also provide analytical expressions for the memory effect associated to GWs. In Sec. \ref{SectionIII}, we modify the formalism, including this time the precession of the orbit in order to evaluate the effects on the emission and on the memory effect. Moreover, we explore the range of validity of the parameters of the system. Finally, we conclude in Sec. \ref{SectionIV}.

\section{Hyperbolic Encounter without orbital precession}
\label{SectionII}
\subsection{Theoretical framework}
\begin{figure}[!t]
\centering
\includegraphics[width = 0.48\textwidth]{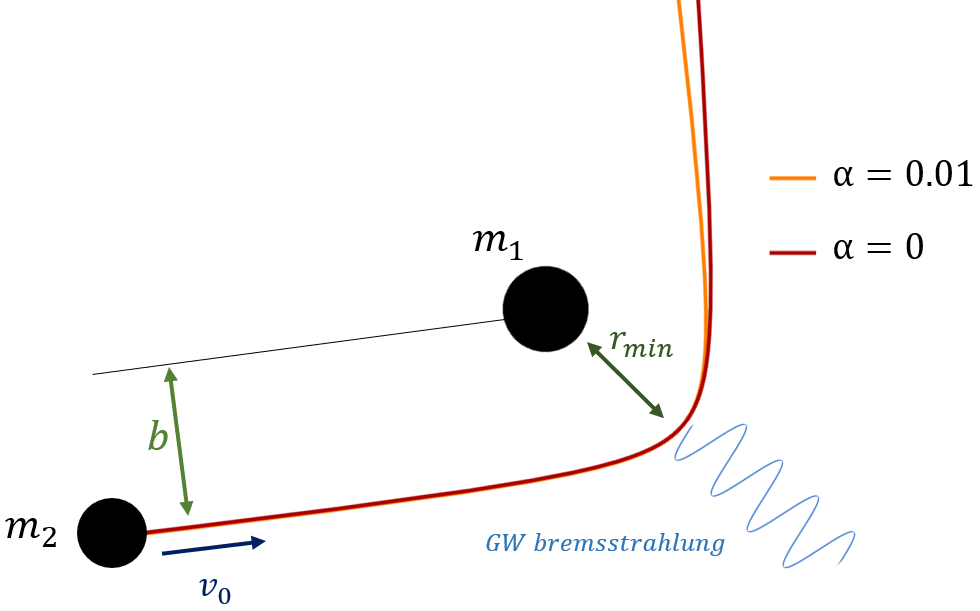}
\caption{Precession of the orbit: representation of the change in hyperbolic orbit of a BH of mass $m_2$ over time, due to the scattering on another of mass $m_1$. This induces the emission of gravitational waves which is maximal at the point of closest approach, $r_{\rm min}$.}
\vspace*{-1mm}
\label{fig:hyperbolic}
\end{figure}

Let us consider a hyperbolic encounter between two compact objects, respectively of mass $m_1$ and $m_2$, where one object is assumed to come from infinity and the other is at rest, without loss of generality. In this framework, the parameters required to model the interaction are the asymptotic velocity (velocity at infinity) $v_0$ and the impact parameter $b$ (see Fig.~\ref{fig:hyperbolic}). 

Moreover, the total mass is given by $M=m_1+m_2$, while the reduced mass is $\mu = m_1 m_2/M$. Knowing the previous quantities, all other parameters to describe the system can be derived. For instance, a first quantity that can be defined is the eccentricity of the hyperbolic orbit, given by
\be\label{eq:eab}
e \equiv \sqrt{1+\frac{b^2}{a^2}} = \sqrt{1+\frac{b^2v_0^4}{G^2M^2}} \ > 1\,,
\ee
where $a$ is the semi-major axis of the orbit and $e$ is the eccentricity of the orbit, which is greater than one for hyperbolic orbits. Decreasing the total mass of the system or increasing the impact parameter or the velocity increases the eccentricity. This also has an impact on GW amplitudes and power spectra of the gravitational radiation emitted, as we will see in what follows. 

We will consider the eccentricity as a given quantity, and we can express the other parameters of the system in terms of $e$. The semi-major axis and the impact parameter turn out to be, respectively,
\begin{align}
 a &= \frac{r_{\rm min}}{e-1}\,, \\
 \label{eq:impact_param}
 b &= r_{\rm min} \sqrt{\frac{e+1}{e-1}}\,.
\end{align}
Equivalently, the maximum approach distance between the two objects during the encounter, $r_{\rm min}$, i.e., the periastron, is given by
\be\label{rmin}
r_{\rm min} = a\,(e-1) = b\,\sqrt{\frac{e-1}{e+1}} \ > R_s \equiv \frac{2GM}{c^2}\,,
\ee
where $R_s$ is the Schwarzschild radius. 
The asymptotic velocity $v_0$ is given by
\be\label{v0}
 v_0 = \sqrt{\frac{(e-1)GM}{r_{\rm min}}} \,,
\ee
and from conservation of angular momentum it results $b\,v_0 = r_{\rm min}\,v_{\rm max}$, where we have to impose that $v_{\rm max} < c$. In addition, we can define the following parameter
\be
\beta \equiv \frac{v_0}{c} < \sqrt{\frac{e-1}{e+1}}\,,
\ee
from which it is possible to derive a lower bound for the impact parameter in terms of the Schwarzschild radius,
\be
b > R_s\,\frac{(e+1)^{3/2}}{2c(e-1)^{1/2}}\,.
\ee
Notice that with the knowledge of the main physical quantities, it is possible to find out the other parameters by inverting the previous formulas and then it is possible to define the orbit and obtain the quadrupole moment of the system, which leads to the calculation of quantities such as GW waveform.

Finally, the orbital trajectory in polar coordinates is given by the usual Keplerian orbit via
\be\label{rphi}
r(\varphi) = 
\frac{a\,(e^2-1)}{1+e\cos(\varphi)}\,.
\ee

\subsection{GW amplitudes}
\begin{figure}[!t]
\centering
\includegraphics[width = 0.48\textwidth]{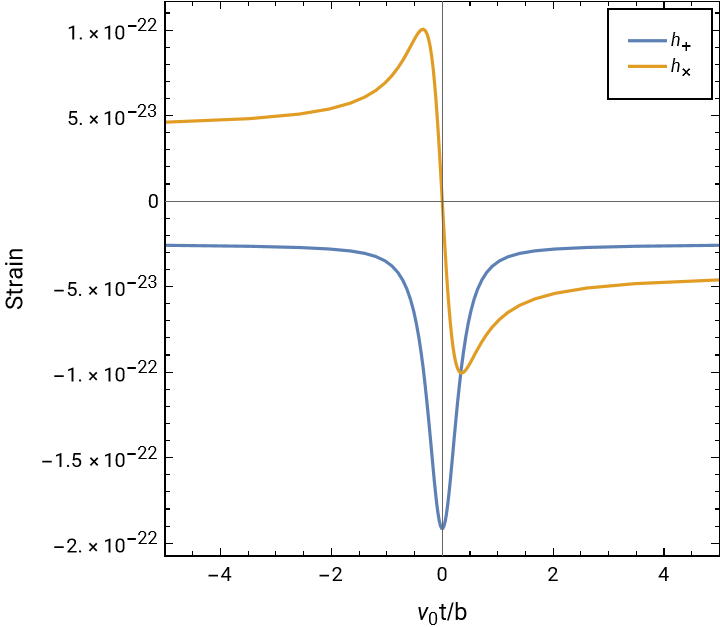}
\caption{Polarization states as functions of the dimensionless combination $v_0 t/b$ for compact objects on hyperbolic orbits.}
\vspace*{-1mm}
\label{fig:polarizations_in_time}
\end{figure}

If we set up a coordinate system such that the position vector $\vec{r}$ is
\be\label{vec_r}
\vec{r} = r(\varphi)\,\big(\cos \varphi,\sin\varphi,0\big),
\ee
then the moment tensor is $M_{ij}=\mu\,r_i\,r_j$ and the reduced quadrupole moment of the system is given by~\cite{Maggiore2007GravitationalWV}
\bea\label{Qij}
Q_{ij} &=& \nn M_{ij}-\frac13 \delta_{ij} M_{kk} \\
&=&\mu \,r^2(\varphi) 
\resizebox{0.7\columnwidth}{!}{
\(
\begin{pmatrix}
\frac16\left(1+3\cos2\varphi\right) & \cos\varphi\sin\varphi & 0 \\
\cos\varphi\sin\varphi & \frac16\left(1-3\cos2\varphi\right) & 0 \\
0 & 0 & -\frac13
\end{pmatrix}
\)},\nn\\
\eea
where $M_{kk}=\mathrm{Tr}(M)$ is the trace of the moment tensor $M_{ij}$. The GW strain amplitude in the TT gauge is given by the second time derivative of the quadrupole moment of the source\cite{Maggiore2007GravitationalWV}
\begin{equation}
 \label{eq:h_ij}
 h_{ij} = \frac{2G}{Rc^4} \stackrel{\cdot\cdot}{Q}_{ij},
\end{equation}
being $R$ the distance from the source to the observer. For simplicity, let us consider the case where the line of sight of the observer is perpendicular to the orbital plane, for instance on $z$ direction. Hence, the two polarization amplitudes turn out to be

\begin{align}
\label{eq:polarizations1}
h_{+} & = -\frac{G\mu v_0^2}{Rc^4(e^2-1)}
\,\Big[ 4 \cos(2 \varphi) \nn \\
&\hspace*{4mm} 
+ e \Big(2e + 5\cos(\varphi) + \cos(3 \varphi)\Big) \Big]\,,
\\
\label{eq:polarizations2}
h_{\times} & = -\frac{G\mu v_0^2}{Rc^4(e^2-1)}
\,\Big[ 4\sin(2\varphi) \nn \\
&\hspace*{4mm} + e \Big(5\sin(\varphi) + \sin(3\varphi)\Big) 
\Big] \,.
\end{align}
These amplitudes are shown in Fig.~\ref{fig:polarizations_in_time}. As an example to obtain a reasonable order of magnitude for GW amplitudes, we consider a hyperbolic encounter of a black hole of mass 30$\,M_\odot$ approaching, at $v_0=0.001\,c$, another black hole of the same mass at rest, with $b\simeq 1.026\,$AU and eccentricity $e=2$. The distance of the system from the observer has been set to $R=30\,$kpc.

At the conclusion of this section, we define the power emitted in GW in the quadrupole approximation, given by~\cite{Maggiore2007GravitationalWV}
\be
\label{eq:powerQij}
  P = \frac{dE}{dt} = \frac{G}{5c^5}\langle \stackrel{\cdots}{Q}_{ij}
  \stackrel{\cdots}{Q}{}^{\!ij}\rangle \, .
\ee
Evaluating the third derivative of the quadrupole moment, the power is equal to~\cite{Garcia-Bellido:2017knh}
\be
P = \frac{32G\mu^2v_0^6}{45c^5\,b^2}\,
f(\varphi,e)\,,
\ee
with
\ba\nonumber
f(\varphi,e) &=& \frac{3\left(1+e\cos(\varphi)\right)^4}
{8(e^2-1)^4}\,\Big[24 + 13e^2 \\[2mm] \label{fephi}
&&\hspace*{-1mm} + 48e\cos(\varphi)
+ 11e^2 \cos(2\varphi)\Big].\,
\ea
Here, $f(\varphi,e)$ is a complicated bell-shaped function of the angle $\varphi$~\cite{Garcia-Bellido:2017knh}.

\subsection{Analysis in the frequency domain}
\begin{figure}[!t]
\centering
\includegraphics[width = 0.48\textwidth]{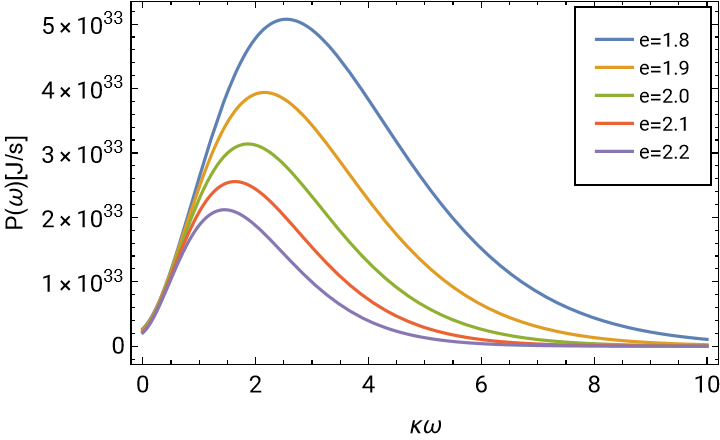}
\caption{Power spectrum as a function of the dimensionless variable $\kappa\omega$ for different values of the eccentricity $e$. The orbital parameters are set to be $m_{1,2}=30\,M_\odot$, $v_0=0.001\,c$, $R=30\,$kpc, while $b$ varies such that we get the desired value of the eccentricity, in particular we find $b_{e=1.8}=0.887\,$AU, $b_{e=1.9}=0.957\,$AU, $b_{e=2.0}=1.026\,$AU, $b_{e=2.1}=1.094\,$AU, $b_{e=2.2}=1.161\,$AU.}
\vspace*{-1mm}
\label{fig:power}
\end{figure}

In the following, we present the result for the power spectrum for the GW emission. The energy released through GWs in the case of hyperbolic encounters between two bodies with masses $m_1$ and $m_2$ is given by~\cite{DeVittori:2012da,Garcia-Bellido:2017knh}
\begin{align}
  \label{eq:DE}
  \Delta E &= \int_{-\infty}^{\infty} P(t)\,dt =
  \frac{8}{15} \frac{G^{7/2}}{c^5}\frac{M^{1/2} m_1^2 m_2^2}{r_{\rm min}^{7/2}}f(e)\,,
\end{align}
where the factor $f(e)$ is given by
\ba
f(e) &=& \frac{1}{(1+e)^{7/2}} \left[24 \arccos \left(-\frac{1}{e}\right)\left(1+\frac{73}{24}e^2+\frac{37}{96}e^4\right) \right.\nn \\ \label{eq:enlossfactor}
&& + \left.\sqrt{e^2-1}\left(\frac{301}{6}+\frac{673}{12}e^2\right)\right]\,.
\ea
The power spectrum can be obtained from the Fourier transform of the energy emission in the time domain,
\be
  \Delta E = \frac{1}{\pi}\int_0^\infty P(\omega)\,d\omega.\label{eq:tot_energy}
\ee
In order to evaluate in Fourier space the power, it is convenient to change variables from the angle $\varphi$ to $\xi$, as the latter characterizes better the hyperbolic orbit and makes the Fourier transform much easier. In this case, the trajectory and the time coordinate are rewritten as \cite{Garcia-Bellido:2017knh}
\ba
r(\xi) &=& a (e\,\cosh\xi-1),\\
t(\xi) &=& \kappa\, (e\,\sinh\xi-\xi), \label{eq:txi}
\ea
with $\kappa=\sqrt{a^3/GM}$, so that the quadrupole tensor becomes
\begin{widetext}
\be
\label{eq:Qij_in_xi}
Q_{ij}=\frac12 a^2 \mu\left(
     \begin{array}{ccc}
      \frac{1}{3} \Big[ (3-e ^2) \cosh 2 \xi -8 e \cosh \xi \Big] &\qquad \sqrt{e ^2-1} (2 e \sinh \xi -\sinh 2 \xi ) &\qquad 0 \\[6pt]
      \sqrt{e ^2-1} (2 e \sinh \xi -\sinh 2 \xi) &\qquad \frac{1}{3}\Big[(2 e ^2-3) \cosh 2 \xi +4 e \cosh \xi\Big] &\qquad 0 \\[6pt]
      0 &\qquad 0 &\qquad \frac{1}{3} \Big[4 e\cosh \xi-e^2 \cosh 2 \xi \Big] \\
     \end{array}
    \right).
\ee
\end{widetext}
Using the Fourier transform (see~\cite{Garcia-Bellido:2017knh} for details), the power in the frequency domain is given by
\be
P(\omega) = \frac{G}{5 c^5}~\sum_{i,j} |\widehat{\dddot{Q}_{ij}}|^2
= \frac{G}{5 c^5}~\omega^6 \sum_{i,j} |\widehat{Q_{ij}}|^2,\label{eq:power}
\ee
where $\widehat{Q_{ij}}$ is the Fourier transform of the quadrupole momentum tensor $Q_{ij}$, which is given in terms of the variable $\xi$ in Eq. (\ref{eq:Qij_in_xi}). Integrating over all frequencies we find the total energy to agree with Eq. (\ref{eq:DE}). In Fig.~\ref{fig:power} the power spectrum is shown for different values of the eccentricity, plotted as a function of the dimensionless variable $\kappa\omega$. Notice that for higher values of eccentricity the peak frequency slowly decreases. This behavior suggests that hyperbolic encounters with lower eccentricities tend to release more energy in the form of GWs, compared to encounters with higher eccentricities.

It should be also noted that these spectra are representative and subject to modification based on the specific values assigned to the parameters. Here, we fix all the parameters to representative values and let the eccentricity vary, for different values of the impact parameter. However, parameters such as masses, relative velocities and impact parameter can determine changes in the eccentricity and therefore, the choice of such values can impact the shape of the spectra.

\subsection{Memory effect}
Furthermore, we have investigated a peculiarity of GW signals, that is the so-called \textit{memory effect}, a property referring to a long timescale difference in the values of the observed metric perturbation associated with the GW~\cite{Favata:2010,Favata:2011}, that is, the late-time and early-time values of at least one of the GW polarizations differ from each other,
\begin{equation}
  \Delta h_{+,\times} = \lim_{t\to +\infty}h_{+,\times} - \lim_{t\to -\infty}h_{+,\times} \neq 0.
\end{equation}

The memory effect in GWs signals is the phenomenon reflecting the displacement of an ideal (truly freely falling) detector after the passing of the wave through it (hence the term \textit{memory}). Such effect is distinguished in a linear and a non-linear type. The linear memory is related to unbounded sources of GWs~\cite{Zeldovich:1974}, such as hyperbolic orbits~\cite{Turner:1977tm, Turner:1978ApJ}, gamma-ray
bursts~\cite{Sago:2004}, dynamical ejection~\cite{Kyutoku:2013}, while the non-linear memory arises from GWs that are sourced by GWs (see, for instance, Refs therein~\cite{Lopez:2023memory}). In particular, since the latter effect derives directly from the radiated GWs and not from the motion of the source, it is present in all sources of GWs, including
bound systems~\cite{Favata:2011}.

While interferometric detectors cannot determine the final amplitude alone as they lack sensitivity at $0$ Hz, for a sufficiently loud source, the frequencies that contribute to the displacement toward the final amplitude are potentially measurable. This could be a tantalizing possibility, especially for future detectors such as LISA and the Einstein Telescope, offering a potential means of extracting valuable information about matter effects from gravitational wave signals emitted by systems of compact objects.

For the hyperbolic case in analysis, at Newtonian order, without considering spin of the massive bodies and without precession of the orbit, we found that only the cross polarization exhibits such effect. Analytically,
\begin{align}
  \Delta h_{+} &= 0 \,,\\
  \Delta h_{\times} &= \frac{8G\mu v_0^2\sqrt{e^2-1}}{Rc^4 e^2}\,.
\end{align}
According to this estimate, there is no linear memory effect in the plus polarization, while there is a difference in the strain of the cross polarization between initial
and final state, depending on the initial system parameter settings.

\section{Hyperbolic encounters with orbital precession}
\label{SectionIII}
\subsection{Theoretical framework \label{sec:theoryPrecession}}
In this section, we want to modify the previous results by considering the precession of the orbit. The precession is a change in the orientation of the orbit due to the gravitational influence of a rotating body around another. To study this phenomenon, we need to describe the evolution of the radial coordinate $r$ as a function of the angular coordinate $\varphi$. In the absence of precession, if an orbit is periodic, for instance in the case of an elliptical orbit, $r(\varphi)$ would be periodic of $2\pi$, denoting that perihelion occurred at the same angular position each orbit~\cite{Carroll}. Instead, in case of precession, the semi-major axis rotates around the central body and the orbits are shifted with respect to each others: this shifting is called \textit{orbital precession}. 

Thanks to perturbation theory, it is possible to show how general relativity (GR) introduces a variation of the period, thus originating the precession~\cite{Carroll}. 
The derivation of the final trajectory is already present in the literature (see, e.g.,~\cite{Carroll,Foster_Nightingale}). In the case of hyperbolic encounters, the solution describing the orbital trajectory which manifests precession of orbit turns out to be
\be\label{eq:r_precession}
r_{\rm pr}(\varphi) = 
\frac{a\,(e^2-1)}{1+e\cos[(1-\alpha)\varphi]}\,,
\ee
where
\be\label{alpha}
\alpha=\frac{3G^2 M^2}{c^2 L^2}
= \frac{3R_s}{2(e+1)r_{\rm min}}.
\ee
Here, the parameter $\alpha$ encodes the effect of the precession of the orbit, with $L$ representing the angular momentum per unit mass, that is $L=b \, v_0$. The orbit alteration due to the precession is graphically reported in Fig.~\ref{fig:hyperbolic}.

It is worth noting that although the factor in Eq. (\ref{alpha}) is small, the effect could become important in some conditions and the corresponding corrections for the measurements might be needed. In particular, if the orbit is highly eccentric, one object can orbit on a ring path around the other one and then escape. Since the emitted GWs carry away energy and angular momentum from the system, this could cause significant deformation in the orbit and also not prevent the capture. The exact outcome of such an encounter depends on several factors, including the masses and spins of the objects and their initial conditions. However, in this limit a different approach is necessary in order to take into account all these factors, such us using higher-order PN corrections or numerical relativity \cite{Jaraba:2021ces}.

It is straight-forward to show that considering a dense BH cluster, as discussed in Ref.~\cite{Trashorras:2020mwn}, there are cases where the effects of the precession cannot be neglected, as the precession parameter becomes larger than $\alpha \ge10^{-4}$. By simulating such dense clusters we see that this is the case in about $\sim2.5\%$ of the cases, thus making it imperative to include this correction in the waveforms.

In what follows we will consider the effect of orbital precession to GW amplitudes and the power spectrum in the frequency domain.
\begin{figure*}[htp]
\centering
{\includegraphics[width=0.46\textwidth]{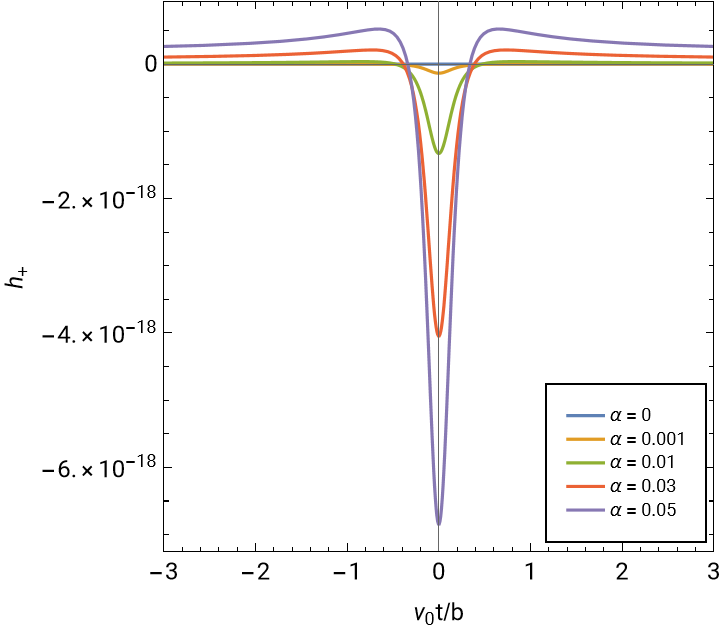}}\qquad\qquad
{\includegraphics[width=0.46\textwidth]{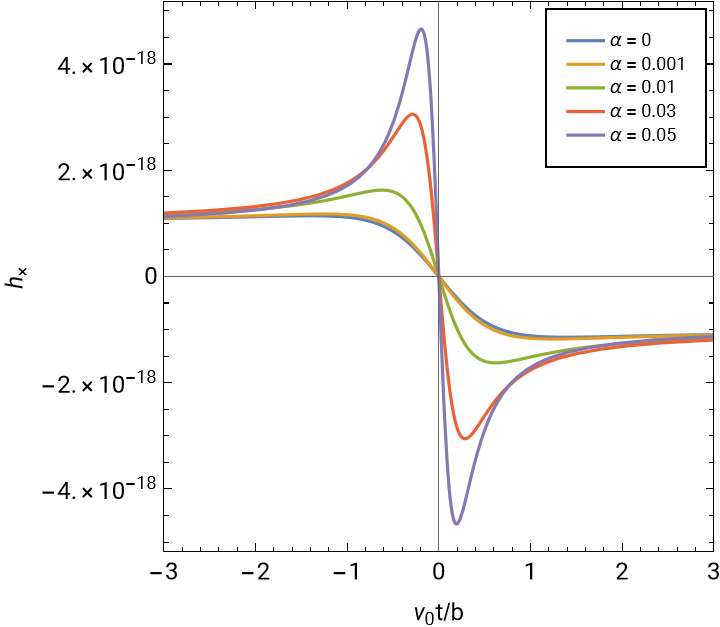}}
\caption{Comparison of $h_{+}$ and $h_\times$, respectively, for different values of $\alpha$, as functions of dimensionless variable $v_0 t/b$. The orbital parameters are set to be $m_{1,2}=30\,M_\odot$, $R=30\,$kpc, while the others varying accordingly. In particular, for $h_{+}$ it results $e_{\alpha=0}\simeq8.102$, $e_{\alpha=0.001}\simeq1.456$, $e_{\alpha=0.01}\simeq1.419$, $e_{\alpha=0.03}\simeq1.416$, $e_{\alpha=0.05}\simeq1.415$.
For $h_\times$ it results $e_{\alpha=0} \gg 1 $, $e_{\alpha=0.001}\simeq31.382$, $e_{\alpha=0.01}\simeq3.548$, $e_{\alpha=0.03}\simeq1.8076$, $e_{\alpha=0.05}\simeq1.504$.
For values $\alpha<10^{-3}$, the effect of precession is very small. To appreciate the shape of the strain in the case $\alpha=0$, especially in the case of the $+$ polarization, see Fig. \ref{fig:polarizations_in_time} as an example. For $\alpha>10^{-2}$ the amplitude of the curves starts to increase and the shapes to be more distorted, but the trend is maintained. Finally, using the fact that the peak frequency is $\omega_{\mathrm{max}}=\frac{v_0}{b}\frac{e+1}{e-1}$, see Ref.~\cite{Garcia-Bellido:2017knh}, we find that for these values of the parameters the peak frequencies are in the range $\omega_{\mathrm{max}}\in[0.006,0.3]\mathrm{Hz}$, so roughly in LISA ranges.}
\label{fig:Plus&CrossPol_in_time}
\end{figure*}
\begin{figure*}[htp]
\centering
{\includegraphics[width=0.46\textwidth]{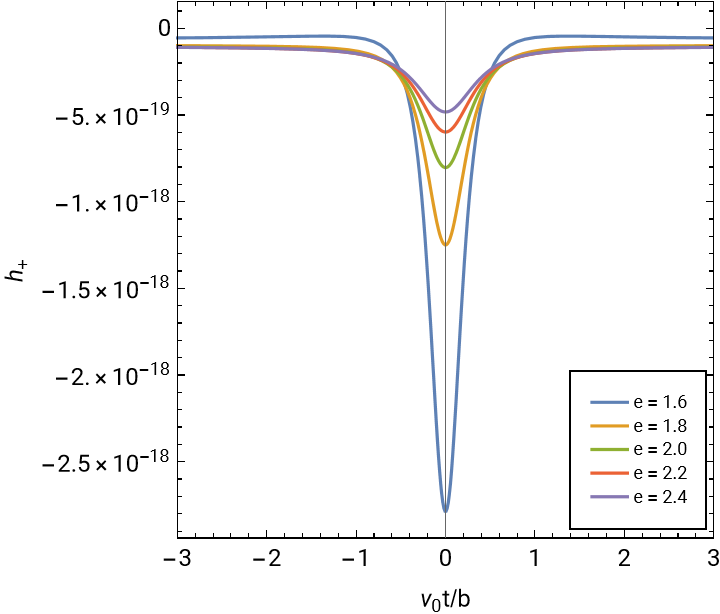}}\qquad\qquad
{\includegraphics[width=0.46\textwidth]{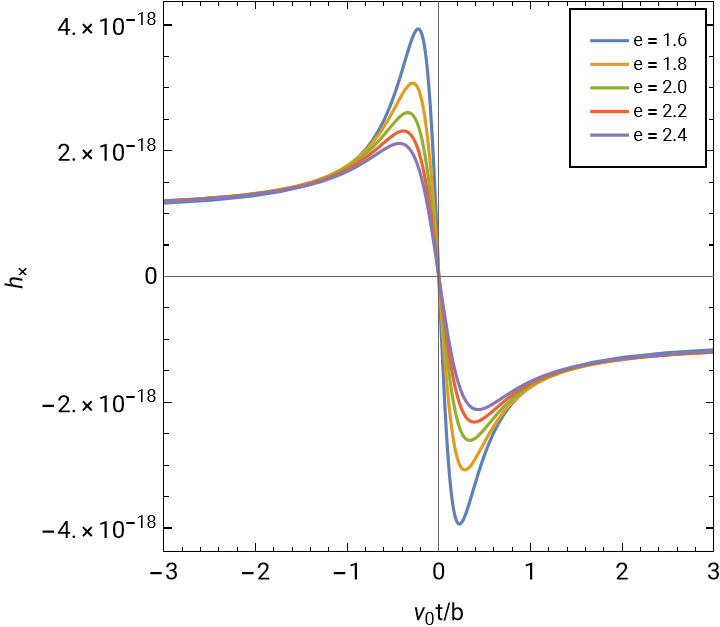}}
\caption{
Comparison of $h_{+}$ and $h_\times$, respectively, for different values of eccentricity $e$, as functions of dimensionless variable $v_0 t/b$. For smaller values of $e$, the amplitudes diverge more and are more peaked. The orbital parameters are set to be $m_{1,2}=30\,M_\odot$, $R=30\,$kpc, while the others varying accordingly. In particular, for $h_{+}$ it results $\alpha_{e=1.6}\simeq1.836\cdot 10^{-2}$, $\alpha_{e=1.8}\simeq7.310\cdot 10^{-3}$, $\alpha_{e=2.0}\simeq4.178\cdot 10^{-3}$, $\alpha_{e=2.2}\simeq2.781\cdot 10^{-3}$, $\alpha_{e=2.4}\simeq2.017\cdot 10^{-3}$.
For $h_\times$ it results $\alpha_{e=1.6}\simeq4.117\cdot 10^{-2}$, $\alpha_{e=1.8}\simeq3.028\cdot 10^{-2}$, $\alpha_{e=2.0}\simeq2.412\cdot 10^{-2}$, $\alpha_{e=2.2}\simeq2.015\cdot 10^{-2}$, $\alpha_{e=2.4}\simeq1.738\cdot 10^{-2}$.}
\label{fig:Plus&CrossPol_diff_ecc}
\end{figure*}

\subsection{GW amplitudes}
In this section, we will present analytic expressions for both GW amplitudes and the comparison with the theoretical predictions in the previous case, without orbital precession.

The GW strain amplitude in terms of quadrupole moment evaluated in the TT gauge is given by Eq. (\ref{eq:h_ij}). Retracing the calculation, the analytical expressions for the polarizations states are given by
\begin{align}
\label{eq:polarizations_alpha1}
h_{+} & = \frac{G\mu v_0^2}{Rc^4(e^2-1)}
\,\Big[ 2 \big(-2 + e^2 (-2 + \alpha) \alpha \big) \cos(2\varphi) \nn \\
&\hspace*{4mm} + e \Big( \big(-1 + (-4 + \alpha) \alpha \big) \cos((-3+\alpha) \varphi) \nn \\
&\hspace*{4mm} - e\,\alpha \cos(2 (-2+\alpha) \varphi) + e (-2+\alpha) \cos(2\alpha\varphi) \nn \\
&\hspace*{4mm} + (-5 + \alpha^2) \cos((1+\alpha)\varphi) \Big) \Big]\,,
\end{align}
\begin{align}
\label{eq:polarizations_alpha2}
h_{\times} & = \frac{G\mu v_0^2}{Rc^4(e^2-1)}
\,\Big[ 2 \big(-2 + e^2 (-2 + \alpha) \alpha \big) \sin(2\varphi) \nn \\
&\hspace*{4mm} + e \Big( \big(1 - (-4 + \alpha) \alpha \big) \sin((-3+\alpha) \varphi) \nn \\
&\hspace*{4mm} + e\,\alpha \sin(2 (-2+\alpha) \varphi) + e (-2+\alpha) \sin(2\alpha\varphi) \nn \\
&\hspace*{4mm} + (-5 + \alpha^2) \sin((1+\alpha)\varphi) \Big) \Big]\,.
\end{align}
Notice that, in the limit of $\alpha\to0$, we recover the previous case, finding that in this case the results are in agreement with Eqs.~(\ref{eq:polarizations1}) and (\ref{eq:polarizations2}). 

The GW polarizations are reported in Fig. 
\ref{fig:Plus&CrossPol_in_time} and Fig.~\ref{fig:Plus&CrossPol_diff_ecc}, respectively for different values of $\alpha$ and $e$. Regarding the plots in Fig.~\ref{fig:Plus&CrossPol_in_time}, it is possible to notice that the curves follow the same trend, especially for lower values of $\alpha$. However, to better appreciate differences in the amplitudes, the reported values of $\alpha$ are quite large. For instance, with the setting of parameters to get GW polarizations in Fig.~\ref{fig:polarizations_in_time}, the value for $\alpha$ turns out to be $\alpha\simeq 10^{-6}$. When $\alpha$ starts approaching larger values, the differences with respect to the case without orbital precession are greater, since the approximations we are working on are broken down and the system is no longer correctly described (see Sec. \ref{validity_range_alpha} for considerations about the validity of our assumptions). A general behavior visible in both panels is that both polarizations are more peaked for larger values of $\alpha$.

In Fig.~\ref{fig:Plus&CrossPol_diff_ecc} we report GWs amplitudes for different values of eccentricity $e$. As the value of $e$ decreases, the amplitudes diverge more and tend to a greater separation between each other. This indicates that lower eccentricity values are associated with stronger GWs bursts.

Note that we present the plots fixing some of the parameters to representative values. This allows us to study the general trend of the amplitudes and understand their behavior. For this purpose, we have considered reasonable values for the eccentricity $e$ and orbital precession $\alpha$, knowing that there exist combinations of values of physical parameters such as mass $M$, impact parameter $b$, and initial velocity $v_0$ that can give those specific values. However, it should be noted that for a comprehensive study, especially for experimental analysis and parameter estimation, it is crucial to determine values of those physical parameters, since the observable is rescaled depending the values of those parameters.

We conclude this subsection by reporting the analytical expressions for the linear memory effect associated to GW signals, as studied in the previous case. Although this difference between asymptotic values in the amplitudes is not visible in the plots, we found that also in this case only the cross polarization exhibits such effect,
\begin{align}
  \Delta h_{+} &= 0 \,,\\
  \Delta h_{\times} &= \frac{2G\mu v_0^2}{rc^4(e^2-1)}
  \Big[ 2\big(-2+e^2(-2+\alpha)\alpha) \sin(2\varphi_0) \nn\\
  &\hspace*{4mm} + e \Big(
  (1-(-4+\alpha)\alpha)
  \sin((-3+\alpha)\varphi_0) \nn\\
  &\hspace*{4mm} + e\alpha\sin(2(-2 +\alpha)\varphi_0)
  + e(-2+\alpha)\sin(2\alpha \varphi_0) \nn\\
  &\hspace*{4mm} + (-5+\alpha^2) \sin((1+\alpha)\varphi_0)
  \Big) \Big]\,,
\end{align}
where
\be
  \varphi_0 = \frac{\arccos (-\frac{1}{e})}{\alpha-1}.
\ee

\subsection{Analysis in the frequency domain}
\begin{figure*}[htp]
 \centering
 {\includegraphics[width = 0.45\textwidth]{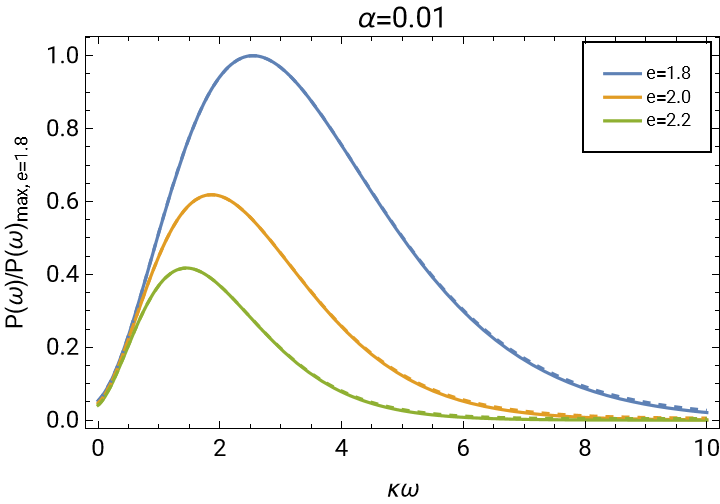}}\qquad\qquad
  {\includegraphics[width = 0.45\textwidth]{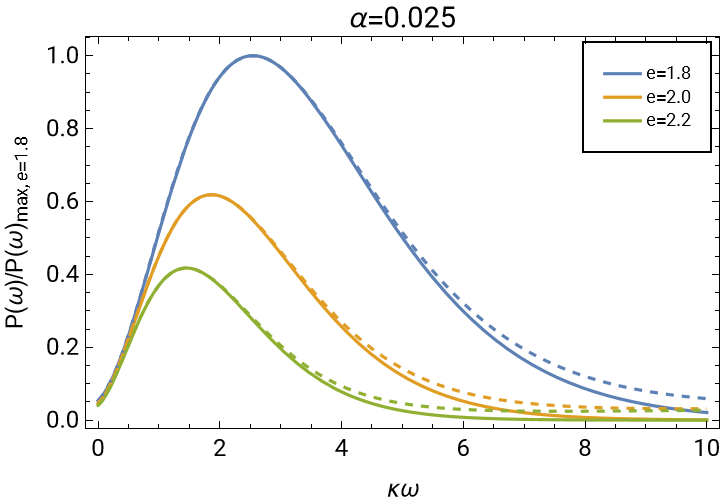}}
  \caption{The GW power spectrum as a function of the dimensionless variable $\kappa\omega$ for different values of eccentricity $e$ and representative values of $\alpha$, respectively, $\alpha=0.01$ and $\alpha=0.025$. The solid lines represent the expected distributions as in the case without orbital precession, while the dashed lines the ones with the $\alpha$ values different from $0$. It is possible to notice, especially for $\alpha=0.025$, that the tails end of the distributions exhibit significant divergences at high frequencies. The maximum power in each curve has been normalized with respect to that of $e=1.8$.}
  \label{fig:Power_alphas}
\end{figure*}
\begin{figure}[!t]
\centering
\includegraphics[width = 0.48\textwidth]{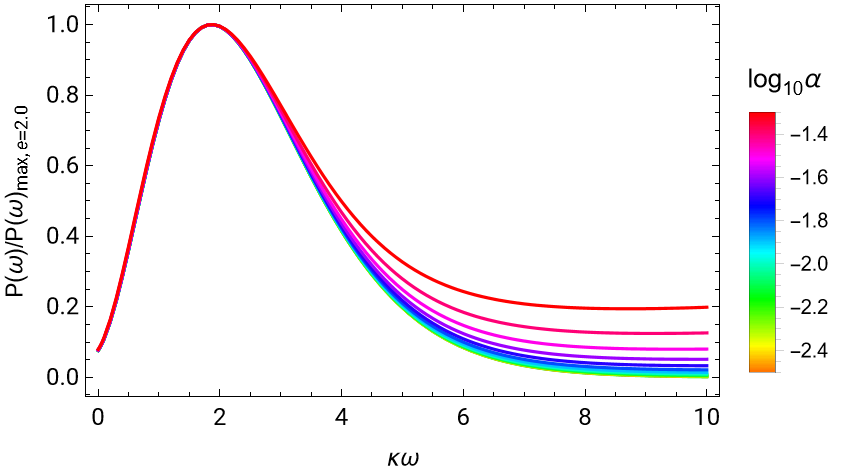}
\caption{The GW power spectrum as a function of the dimensionless variable $\kappa\omega$ for different values of $\alpha$, in the case of $e=2$. As $\alpha$ is increased, the tails of the power spectra undergo considerable amplification and start to diverge further at high frequencies. This behavior indicates that the assumption of a precessing semi-Keplerian orbit can no longer be considered sufficient to describe the system.}
  \label{fig:Power_alphas_log}
\end{figure}
\begin{figure*}[htp]
\centering
{\includegraphics[scale=0.46]{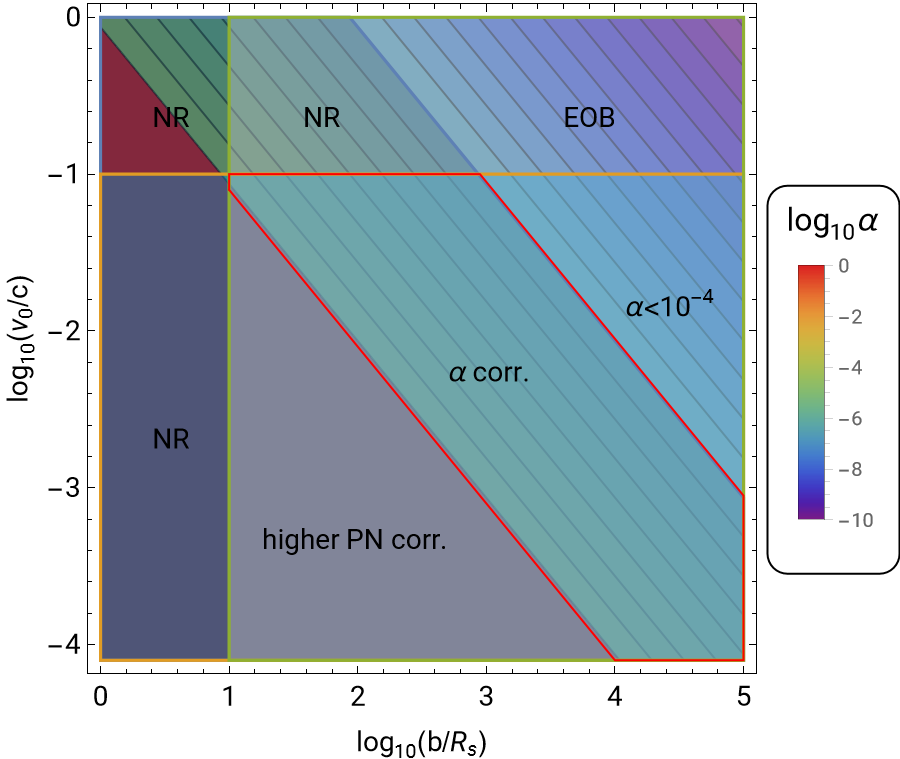}}\qquad\qquad
{\includegraphics[scale=0.5]{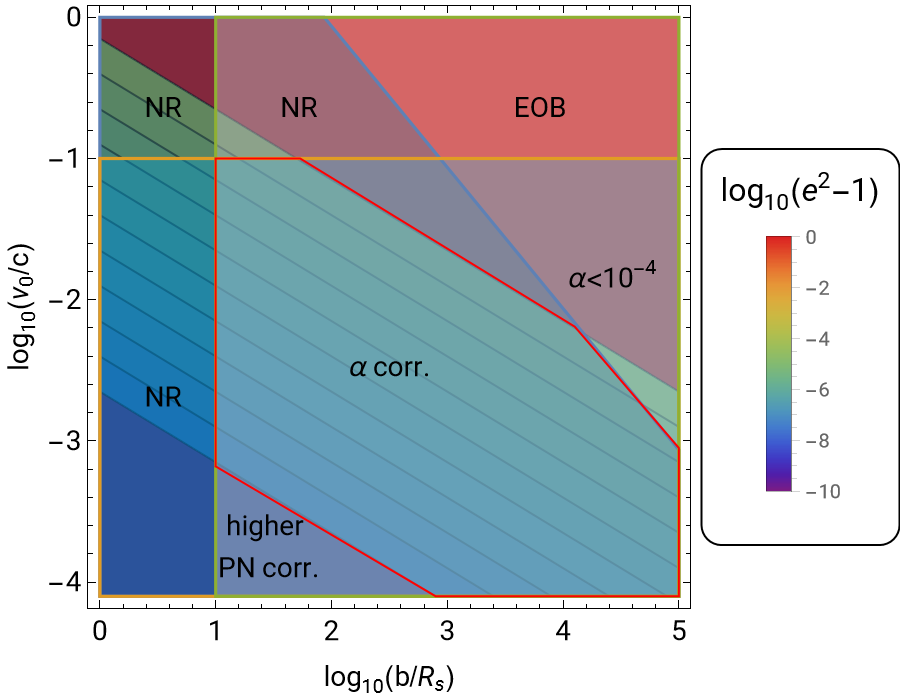}}
\caption{Contour plots of the parameter space that highlight the various regimes of the CHE. The $x$-axis represents the impact parameter normalized by the Schwarzschild radius $R_s$, while the $y$-axis the asymptotic velocity normalized by the speed of light, in logarithmic scale. Here, \textit{NR} stands for Numerical Relativity and outlines the region where it is needed for a proper analysis of the system. \textit{EOB} stands for effective-one-body theory, which is one of the possible methods to describe high-velocity, small-gravity regime. On the other hands, \textit{higher PN corr.} indicates the parameter region where using the PN expansion method is required, including higher order corrections. Finally, \textit{$\alpha$ corr.} delimits the area where our analysis with the modification of the trajectory is valid. Note that it is still valid also for values of $\alpha<10^{-4}$, but the differences in the observables would not be appreciable, as already pointed out.
The contour lines delineate the range of parameter values for which the analysis of hyperbolic encounters is well defined. The left panel defines the validity area by varying $\alpha$, while the right one by varying the eccentricity $e$.}
\label{fig:range_alpha}
\end{figure*}
By setting the system as in the previous case, we start from the position vector $\vec{r}$ as in Eq. (\ref{vec_r}), where, this time, we have to substitute the definition of $r_{\rm pr}$ given by Eq. (\ref{eq:r_precession}). Then, the reduced quadrupole moment of the system is formally given by the same definition of Eq. (\ref{Qij}), but with the substitution of $r_{\rm pr}$. 
Retracing the same calculations as in the case without precession, the power emitted in GW is again given by the expression contained in Eq. (\ref{eq:powerQij}), but this time the quadrupole moment turns out to be so complicated and not linear that it cannot be treated analytically as in the case without precession. 

For this reason, we need to numerically integrate and then get the energy variation in terms of the $\alpha$ parameter and the eccentricity $e$ (see Appendix \ref{sec:extra1} for details). This trend is shown in Fig.~\ref{fig:Power_alphas}. We report the power spectrum as a function of the dimensionless variable $\kappa\omega$, comparing two different cases, i.e., for $\alpha=0.01$ and $\alpha=0.025$. In each plot, the solid line represents the case without precession, so it is possible to note that we also numerically cover the previously obtained results. The dashed line represents the comparison with the $\alpha$ value different from $0$. For increasing values of the $\alpha$ parameter, the divergences between curves start to be consistent, as we are approaching limit values for the validity of our assumptions. Notice that we have selected specific values for the eccentricity $e$ and orbital precession parameter $\alpha$ by setting all the other parameters to representative values. However, as already pointed out, certain combinations of physical parameters can yield the specified values for $e$ and $\alpha$.

In Fig.~\ref{fig:Power_alphas_log}, we report the power spectra by exploring different values of $\alpha$. Again, the values of all parameters have been set to representative values in order to obtain the shape such that $e$ remains fixed while $\alpha$ varies according to the values shown in the plot. The differences in the tails of the curves are due to the fact that the system is no longer properly described with our assumption of a precessing semi-Keplerian orbit. The significant divergences are therefore only due to the fact that our description of the system breaks down at some value of $\alpha$ at these frequencies, thus it is necessary to include higher order PN corrections in order to obtain an appropriate and no longer divergent curve behaviour. 

In fact, we can actually estimate precisely the values of $\alpha$ for which our description breaks down. One indicator for the latter is when the total energy emitted by the system no longer is finite, as the integral of Eq.~\eqref{eq:tot_energy} diverges. In the non-precessing case it was shown in Ref.~\cite{Garcia-Bellido:2017knh} that the power as a function of frequency falls approximately exponentially as a function of frequency, i.e. $\sim e^{-n_0\kappa\, \omega}$ for some real positive number $n_0$, thus ensuring the convergence of the integral. Thus, in this case, we fit the tails of the curves of Fig.~\ref{fig:Power_alphas_log} to exponentials of the form $P(\kappa \, \omega\gg 1)\approx e^{-n\,\kappa\, \omega}$ and we find that for $\log_{10}\alpha\simeq -1.51$ or $\alpha \simeq 0.031$, the variable $n$ changes sign from negative (for $\alpha < 0.031$) to being positive, thus making the integral of Eq.~\eqref{eq:tot_energy} diverging.

Finally, as one can notice also in this case, even in the zero frequency limit, there is energy emitted by the system~\cite{Garcia-Bellido:2017knh,DeVittori:2014psa}. This is due to the fact that cross polarization exhibits the memory effect.

\subsection{Validity range of parameters}
\label{validity_range_alpha}
In the following, we want to comment about the validity of our assumptions and, therefore, on reasonable values of the parameters of the system, in particular $\alpha$ and $e$.

Let us remember the definition of $\alpha$ in terms of the eccentricity, that is Eq. (\ref{alpha}). In this case, it is possible to notice that for $e\simeq 1$, which corresponds to the parabolic trajectory, the $\alpha$ parameter tends to a constant values, while, for high values of $e$, that is for high values of the asymptotic velocity, $\alpha\to 0$. The latter can be explained as follows: for this values of eccentricity no curve is defined, but a line. This means that, due to the imposed conditions, one body moves so fast that it does not affect each other with the other body at rest and goes straight for its trajectory.

In Fig.~\ref{fig:range_alpha}, we report the contour plots for the validity region of our approximation. However, it has to be kept in mind that constraints on values of $\alpha$ and $e$ are due to those imposed on the physical quantities of interest, such as the asymptotic velocity and maximum encounter distance. In particular, we do not consider values less than $\alpha=10^{-4}$, as differences with respect to the case without orbital precession would not be appreciable. 

On $x$-axis we consider the impact parameter $b$ normalized by the Schwarzschild radius. For values of $b$ approaching the Schwarzschild radius our treatment in terms of $\alpha$ parameter is no longer correct and numerical relativity is necessary to describe the interaction.

On the $y$-axis we consider the asymptotic velocity $v_0$ normalized by the speed of light. Consistent values for the validity of this approach are $v_0\simeq 10^{-2} c$: for higher velocities numerical relativity is again necessary.

In the left panel of Fig.~\ref{fig:range_alpha}, we vary the $\alpha$ parameter, selecting as the contour region the one for $10^{-4}<\alpha<10^{-2}$. It is worth noting again that our description by directly modifying the trajectory is still valid for values of $\alpha<10^{-4}$. However, for such values of the parameter $\alpha$ the differences on the observables would not be appreciable. By varying $\alpha$ we are essentially changing the ratio of the Schwarzschild radius to the minimum separation distance, see Eq.~\eqref{alpha} and remember that $b$ and $r_{\rm min}$ are related by Eq.~\eqref{eq:impact_param}. The Schwarzschild radius represents the gravitational influence of a compact object and is directly related to its mass. The minimum separation distance, on the other hand, reflects the closest approach of the two objects during the hyperbolic encounter.

In the right panel, we are varying the eccentricity of the system. The selected contour region encodes values of $1<e< 2\times 10^5$, with the maximum value corresponding to the extreme values of the parameter space of the contour plot, i.e $b/R_s=10^5$ and $v_0/c=1$. By varying $e$, we are in fact varying the particular shape of the orbit.

Let us conclude this section with a remark on the eccentricity and semi-major axis. As said, during a hyperbolic encounter between massive compact objects, GWs are emitted as a result of the objects interaction and motion. These GWs carry away energy and angular momentum from the system, causing changes in the orbital parameters, in particular the eccentricity $e$ and semi-major axis $a$. In the context of Newtonian theory, these quantities are constants of motion. However, in GR they become time-dependent functions, being related to the total energy and relative angular momentum~\cite{Peters:1964}. This change might be negligible compared to the initial values for $e$ and $a$. Therefore, by fixing them to specific values in our analysis, we are assuming that the emitted GWs have a minimal impact on altering the eccentricity and semi-major axis of the orbit during the encounter. This assumption allows us to facilitate the analysis. 

\begin{figure*}[!tp]
\centering
\includegraphics[width = 0.48\textwidth]{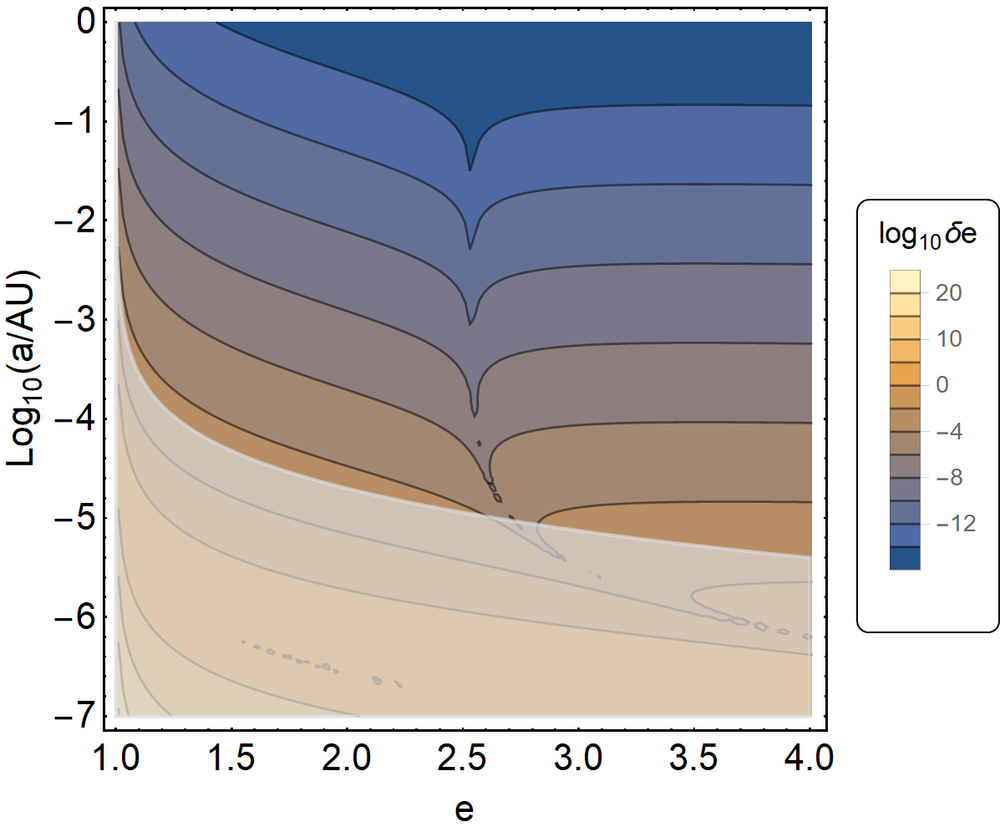}\qquad
\includegraphics[width = 0.48\textwidth]{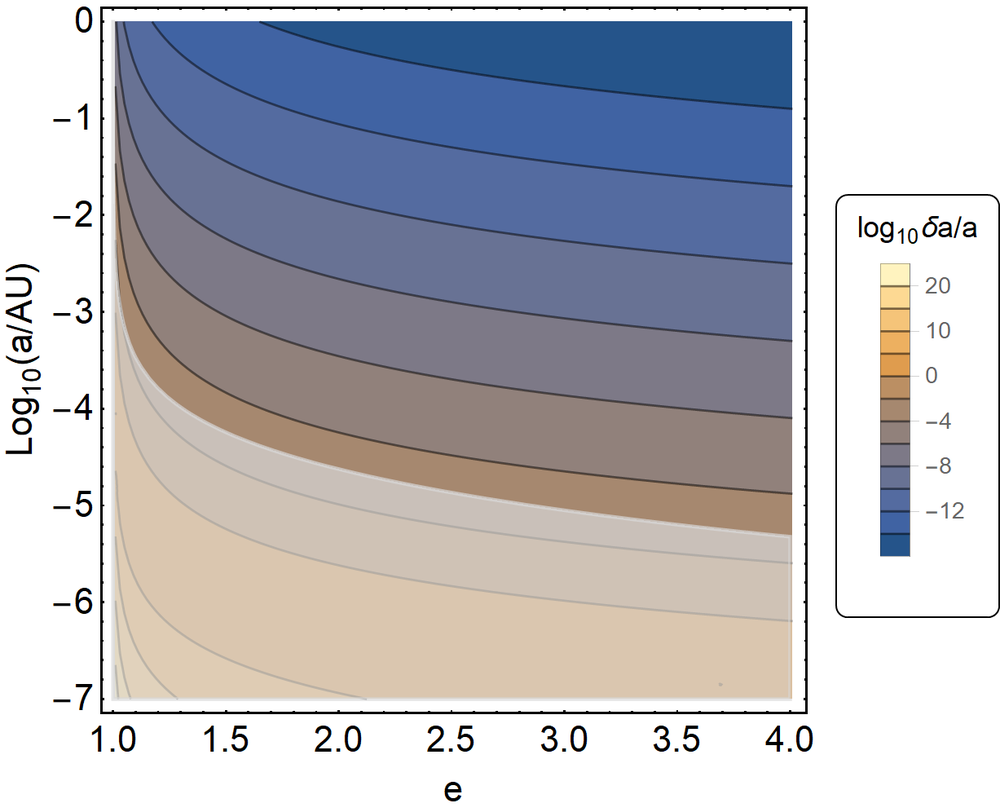}
\caption{Contour plots of the variation $\delta e$ of the eccentricity and $\frac{\delta a}{a}$ of the semi-major axis, taking into account the energy and angular momentum loss, for two black holes of masses $m_{1,2}=30\,M_\odot$, using Eqs.~\eqref{eq:de} and \eqref{eq:da_a}. The gray-shaded region corresponds to the approximate region where the precession approximation completely breaks down (roughly $\alpha\sim 0.03$). \label{fig:da_de_contour}}
\end{figure*}

This aspect was already pointed out in Ref.~\cite{Garcia-Bellido:2017knh} for the simplest case without considering orbital precession. In order to be sure of the validity of this assumption, we evaluate the energy radiated $E$ from this system by GWs and the angular momentum $L$~\cite{Peters:1964} and then perform the variation in time of the eccentricity and the semi-major axis in the case of hyperbolic orbit
\ba
  \delta e &\simeq& \frac{L(L\,\Delta E + 2 E \Delta L)}{G^2\,M^2\,e\,\mu^3} \,,\label{eq:de}\\
  \frac{\delta a}{a} &\simeq& -\frac{G\,M\,\mu \Delta E}{2E^2} \,.\label{eq:da_a}
\ea
After these evaluations, we find that such variations are negligible for realistic values. For instance, in the case of a hyperbolic encounter between two black holes of 30 $M_\odot$, with $a=0.01$ AU, $\alpha\sim 10^{-4}$ and eccentricity $e=2$, we get both variations of the order of $10^{-10}$. 

Also, in Fig.~\ref{fig:da_de_contour} we show contour plots of the variation $\delta e$ of the eccentricity and $\frac{\delta a}{a}$ of the semi-major axis, taking into account the energy and angular momentum loss, for two black holes of masses $m_{1,2}=30\,M_\odot$. As can be seen in both panels of Fig.~\ref{fig:da_de_contour}, for reasonable values of the semi-major axis and the eccentricity, the variation in the orbital elements of the hyperbolic orbit is vanishingly small and only become important when the semi-major axis becomes too small (below $10^{-5}$ AU). Therefore, we find that it is possible to keep the value of the eccentricity and semi-major axis fixed during all the analysis of the emitted GW signal.

Finally, we note that there is a ``glitch" on the left panel of Fig.~\ref{fig:da_de_contour}, due to the way we plot the contours, namely the function changes sign at $e\sim 2.5$ for large values of the semi-major axis $a$, while as the two bodies get closer, relativistic effects kick-in and push the sign change to higher eccentricities. For $\alpha=0$ we can calculate exactly where this change of sign happens using Eq.~\eqref{eq:de} and expanding the expression. Doing so we find:
\ba 
&&\delta e(e, \alpha =0) \nn\\
&=& \frac{2 G^{5/2} \mu\,m^{3/2}}{45 a^{5/2}\,c^5}\,\frac{1}{e \left(e^2-1\right)^3}\,\Bigg[\nn 72 e^6-349 e^4-793 e^2\\
& + &3 \left(47 e^4-280 e^2-192\right) \sqrt{e^2-1} \sec ^{-1}(-e)+1070\Bigg],\nn \\\label{eq:de_func}
\ea
which we can easily see, by solving numerically the equation $\delta e(e, \alpha =0)=0$, that it flips sign at $e\simeq 2.536$. Doing a similar analysis including the effects of precession results in an expression for $\delta e$ that, at linear order, depends on $\alpha$ as:
\be 
\delta e(e, \alpha) = \delta e(e, \alpha =0)+ \delta e_1(e, \alpha =0)\,\alpha+\mathcal{O}(\alpha^2),
\ee 
where $\delta e_1(e, \alpha =0)$ is a complicated expression that depends only on the eccentricity $e$. This then causes a shift of the root of $\delta e(e, \alpha)=0$ to the right in the left panel of Fig.~\ref{fig:da_de_contour}.

\section{Conclusions}
\label{SectionIV}
In this paper, we have offered a comprehensive theoretical overview about hyperbolic encounters between massive compact objects, including the effects of orbital precession. These encounters can occur in dense clusters where two objects interact gravitationally and undergo a scattering event. During the closest approach of the objects, they emit Bremsstrahlung gravitational radiation, which has the potential to be detected by future interferometers.

This analysis has involved non-spinning binaries and the calculations have been performed by including the effects of precession via the $\alpha$ parameter in the orbital equation, as in Eq.~\eqref{eq:r_precession}. The aim was to offer an initial estimate of the effects caused by the precession of the orbit due to the gravitational interaction between two objects. In particular, the waveform resulting from a hyperbolic encounter differs significantly from the waveform produced by a binary inspiral. The burst-like structure of the signal in hyperbolic encounters carries distinctive features that can be used in parameter estimation to extract crucial information about the system.

We have presented a review of the theory underlying hyperbolic encounters, at first excluding the effects of orbital precession. We have derived the main relations governing the geometry and physics of these encounters and provided analytical expressions for the strain amplitude and power spectrum of the emitted GWs. Thereafter, we have extended the formalism to include orbital precession and evaluate its effects on the emission. We have directly modified the trajectory expression, by incorporating in the $\alpha$ parameter some general relativistic effects between the objects responsible for precession. It is worth noticing that although the effect of orbital precession on GW signals is small, it properly accounts for relativistic effects in the system. Additionally, we have derived analytical expressions for the linear memory effect associated with GWs in both cases, finding that only the cross polarization state exhibits such effect. This means that there is a non-vanishing difference between the amplitude of the signal at early and late times. Finally, we have explored the range of validity for the parameters of the system and discussed the implications of our results. 

In conclusion, hyperbolic encounters between massive compact objects and their GW signatures could provide valuable information that can help in estimating parameters and broaden our knowledge of these intriguing phenomena. This is also a challenge from an experimental point of view, as in the future it will be necessary to disentangle these signals from typical interferometer noise bursts. Detection and analysis of these events would complement observations of binary mergers in the observable Universe and offer insights into the nature of the objects that originated them.

\section*{Acknowledgements}
The authors acknowledge support from the research project PID2021-123012NB-C43 and the Spanish Research Agency (Agencia
Estatal de Investigaci\'on) through the Grant IFT Centro de Excelencia Severo Ochoa No CEX2020-001007-S, funded by MCIN/AEI/10.13039/501100011033. M.C. acknowledges support from a SO PhD fellowship. S.K. is supported by the Spanish Atracci\'on de Talento contract no. 2019-T1/TIC-13177 granted by Comunidad de Madrid, the I+D grant PID2020-118159GA-C42 funded by MCIN/AEI/10.13039/501100011033, the i-LINK 2021 grant LINKA20416 of CSIC, and Japan Society for the Promotion of Science (JSPS) KAKENHI Grant no. 20H01899, 20H05853, and 23H00110.

\appendix

\section{Numerical integration of the Fourier transforms related to the power emitted \label{sec:extra1}}

Here we briefly discuss the numerical integration techniques required to perform the Fourier transforms of the quadrupole tensor in the case of precessing orbits, as then it is not straight-forward to perform the calculations analytically in terms of Hankel functions, as in the non-precessing case.

The problem arises as in the case of hyperbolic encounters for the Fourier transforms we need to numerically integrate all the way to infinity rapidly oscillating functions that are not bounded at infinity, eg $\cosh\xi$ etc, but also very complicated forms in terms of the precession parameter $\alpha$. 

To avoid this issue, the following technique is implemented. First, to avoid numerical issues from the numerical integration, we break the integration of the Fourier transform into parts of length equal to the period $\omega$ and then perform the infinite, but converging, sum using the command \texttt{NSum} in \textit{Mathematica}. For example, the Fourier integral of the quadrupole tensor becomes 
\ba 
\widehat{Q}_{ij}(\alpha,\omega)= \sum_{j=-\infty}^{\infty} \int_{j\, \pi/\omega}^{(j+1)\, \pi/\omega} e^{i\, \omega\,t(\xi)}\,Q_{ij}(\alpha,\xi)\, dt,~~~~
\ea 
where $Q_{ij}(\alpha,\xi)$ is the quadrupole tensor including precession and $t(\xi)$ is given by Eq.~\eqref{eq:txi}.

Second, we perform a linear expansion of the quadrupole tensor in terms of $\alpha$, in order to speed up the calculations by removing the dependence on this parameter. This approximation is valid in the context of our analysis, as we expect the system to be non-relativistic, i.e. $\alpha\ll 1$. So, we expand the quadrupole tensor as
\be 
Q_{ij}(\alpha,\xi)\simeq Q_{ij}(\alpha=0,\xi)+\alpha\, \delta Q_{ij}(\xi)+\ldots,
\ee 
where $\delta Q_{ij}(\xi)$ is a complicated expression resulting from the Taylor expansion, while in this case $Q_{ij}(\alpha=0,\xi)$ is given by Eq.~\eqref{eq:Qij_in_xi}. Doing so has the advantage that we can calculate the Fourier transforms of $Q_{ij}(\alpha=0,\xi)$ and $\delta Q_{ij}(\xi)$ only once for every frequency $\omega$, without having to take into account $\alpha$, thus significantly speeding up the calculations.

Therefore, we can write the Fourier transformed quadrupole tensor as 
\be 
\widehat{Q}_{ij}(\alpha,\omega)\simeq \widehat{Q}_{ij}(\alpha=0,\omega)+\alpha\, \widehat{\delta Q_{ij}}(\omega)+\ldots,
\ee 
and then the power can be shown to be, using Eq.~\eqref{eq:power}, quadratic in the precession parameter $\alpha$.

To validate our methodology and make sure any numerical errors are under control, we compare the results of the analysis with those of the case without precession ($\alpha=0$), where we know the exact analytic result in terms of Hankel functions. In fact, we find excellent agreement between the two methods in the $\alpha=0$ limit, thus confirming our analysis.

%\clearpage
\bibliography{GWbursts}

\end{document}